\documentclass[runningheads,a4paper]{llncs}

\usepackage{pbox}
\usepackage{diagbox}
\usepackage{amssymb}
\usepackage{graphicx}
\usepackage{epsfig}
\usepackage{amsmath}
\usepackage{amsfonts}
\usepackage{arydshln}
\usepackage{url}
\usepackage{graphicx}
\usepackage{lscape}

\usepackage{csquotes}

\usepackage{pbox}
\usepackage{amssymb}
\usepackage{graphicx}
\usepackage{epsfig}
\usepackage{amsmath}
\usepackage{amsfonts}
\usepackage{arydshln}
\usepackage{url}
\usepackage{graphicx}
\graphicspath{{poster_graphics/}{graphics/}{paper_graphics/}{New_plots/}}
\usepackage{lipsum}
\usepackage{epsfig}
\usepackage{graphicx,epstopdf}
\DeclareGraphicsExtensions{.ps}
\usepackage{verbatim}
\usepackage{multicol}
\usepackage{subfigure}

\urldef{\mailsa}\path|{alfred.hofmann, ursula.barth, ingrid.haas, frank.holzwarth,|
\urldef{\mailsb}\path|anna.kramer, leonie.kunz, christine.reiss, nicole.sator,|
\urldef{\mailsc}\path|erika.siebert-cole, peter.strasser, lncs}@springer.com|    
\newcommand{\keywords}[1]{\par\addvspace\baselineskip
\noindent\keywordname\enspace\ignorespaces#1}
\begin{document}

\mainmatter  

\title{It's Only Words And Words Are All I Have\footnote{A line from song Words by Bee Gees}}

\titlerunning{It's Only Words And Words Are All I Have}

%
%
\author{Manash Pratim Barman\inst{1} \and
Kavish Dahekar\inst{2} \and
Abhinav Anshuman\inst{3}\and 
Amit Awekar\inst{4}}
\authorrunning{M. P. Barman et al.}
%

\institute{Indian Institute of Information Technology, Guwahati \and
SAP Labs, Bengaluru \and
Dell India R\&D Center, Bengaluru \and 
Indian Institute of Technology, Guwahati
\\
\email{awekar@iitg.ac.in}}
%
%

\toctitle{It's Only Words And Words Are All I Have}
\tocauthor{Manash Pratim Barman, Kavish Dahekar, Abhinav Anshuman and Amit Awekar}
\maketitle

\begin{abstract}

The central idea of this paper is to demonstrate the strength of lyrics for music mining and natural language processing (NLP) tasks using the distributed representation paradigm. For music mining, we address two prediction tasks for songs: genre and popularity. Existing works for both these problems have two major bottlenecks. First, they represent lyrics using handcrafted features that require intricate knowledge of language and music. Second, they consider lyrics as a weak indicator of genre and popularity. We overcome both the bottlenecks by representing lyrics using distributed representation. In our work, genre identification is a multi-class classification task whereas popularity prediction is a binary classification task. We achieve an F1 score of around 0.6 for both the tasks using only lyrics. Distributed representation of words is now heavily used for various NLP algorithms. We show that lyrics can be used to improve the quality of this representation.

\end{abstract}

%




\section{Introduction}


The dramatic growth in streaming music consumption in the past few years has fueled the research in music mining~\cite{nielson17}. More than 85\% of online music subscribers search for lyrics~\cite{midia17}. It indicates that lyrics are an important part of the musical experience. This work is motivated by the observation that lyrics are not yet used to their true potential for understanding music and language computationally. There are three main components to experiencing a song: visual through video, auditory though music, and linguistic through lyrics. As compared to video and audio components, lyrics have two main advantages when it comes to analyzing songs. First, the purpose of the song is mainly conveyed through the lyrics. Second, lyrics as a text data require far fewer resources to analyze computationally. In this paper, we focus on lyrics to demonstrate their value for two broad domains: music mining and NLP. 
 
 \begin{figure}[ht]
\centering
\begin{minipage}{0.65\textwidth}
  \centering
  \includegraphics[height=1.2in, width=2.5in]{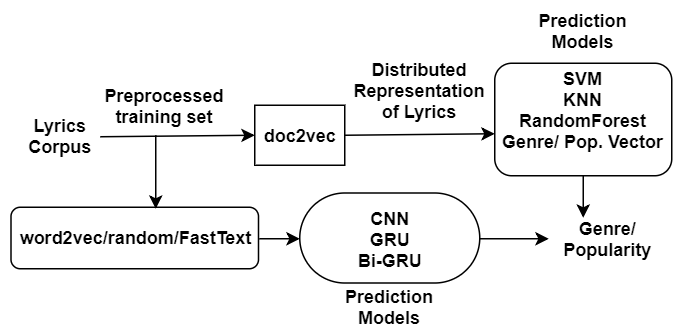}
  \caption{Genre \& Popularity Prediction}
  \label{fig1}
\end{minipage}%
\begin{minipage}{0.3\textwidth}
  \centering
  \includegraphics[height=1.2in, width=1in]{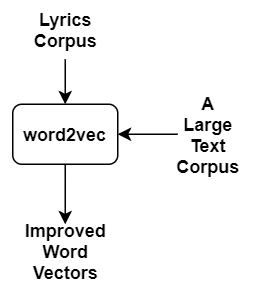}
  \caption{Improving Word Vectors}
  \label{fig2}
\end{minipage}
\end{figure}

A recent trend in NLP is to move away from handcrafted features in favor of distributed representation. Methods such as word2vec and doc2vec have achieved tremendous success for various NLP tasks in conjunction with Deep Learning~\cite{DBLP:journals/cim/YoungHPC18}. Given a song, we focus on two prediction tasks: genre and popularity. We apply distributed representation learning methods to jointly learn the representation of lyrics as well as genre \& popularity labels. Using these learned vectors, we experiment with various traditional supervised machine learning and Deep Learning models. We apply the same methodology for popularity prediction. Please refer to Figures~\ref{fig1} and ~\ref{fig2} for overview of our approach. 

Our work has three research contributions. First, this is the first work that demonstrates the strength of distributed representation of lyrics for music mining and NLP tasks. Second, contrary to existing work, we show that lyrics alone can be good indicators of genre and popularity. Third, the quality of words vectors can be improved by capitalizing on knowledge encoded in lyrics.

\section{Dataset}

Lyrics are protected by copyright and cannot be shared directly. Most researchers in the past have used either small datasets that are manually curated or large datasets that represent lyrics as a bag of words~\cite{CAMP12,RudolfACM,LoganIEEE,6204985,Hu:2012:GCM:2348283.2348480,McKay2010EvaluatingTG,Fell2014LyricsbasedAA}. Small datasets are not enough for training distributed representation. Bag of words representation lacks information about the order of words in lyrics. Such datasets cannot be used for training distributed representation. To get around this problem, we harvested lyrics from user-generated content on the Web. Our dataset contains around 400,000 songs in English. We had to do extensive preprocessing to remove text that is not part of lyrics. We also had to detect and remove duplicate lyrics. Metadata about lyrics that is genre and popularity was obtained from Fell and Sporleder \cite{Fell2014LyricsbasedAA}. However, for genre and popularity prediction, we were constrained to use only a subset of dataset due to class imbalance problem.

\begin{table*}[t]
\centering
\caption{F1-Scores for Genre Prediction. Highest value for each genre is in bold.}
\label{table:genre}
\begin{tabular}{|c|c|c|c|c|c|c|c|c|c|}
\hline
  \backslashbox{ Model }{ Genre }            & Metal          & Country        & Religious      & Rap            & R\&B          & Reggae         & Folk           & Blues         & Average        \\
\hline

SVM             & 0.575          & 0.493           & 0.634         & \textbf{0.815} & 0.534         & 0.608          & 0.437           & 0.532         & 0.579          \\
\hline

KNN             &0.463          & 0.457          & 0.557          & 0.729          & 0.457         & 0.547          & 0.428          & 0.515         & 0.519          \\
\hline

Random Forest        & 0.552          & 0.536         & 0.644          & 0.791  & 0.525         & 0.599          & 0.474          & 0.559         & 0.585           \\
\hline

Genre Vector   & \textbf{0.605} & \textbf{0.551} & 0.641 & 0.738          & \textbf{0.541} & \textbf{0.716} & \textbf{0.475} & \textbf{0.59} & \textbf{0.607} \\
\hline

CNN   & 0.543 & 0.466 & \textbf{0.668} & 0.801          & 0.504 & 0.628 &0.471 & 0.563 & 0.580 \\
\hline

GRU   & 0.479 &   0.467 & 0.558          & 0.745 & 0.462 & 0.601 & 0.355 & 0.531 & 0.525 \\
\hline

Bi-GRU & 0.494 & 0.471 & 0.567 & 0.752         & 0.492 & 0.609 & 0.372 & 0.488 & 0.531 \\
\hline

\end{tabular}
\end{table*}

\section{Genre Prediction}
\label{section:genre}
Our dataset contains songs from eight genres: Metal, Country, Religious, Rap, R\&B, Reggae, Folk, and Blues. Our dataset had a severe problem of class imbalance with genres such as Rap dominating. Using complete dataset was resulting in prediction models that were highly biased towards the dominant classes. Hence, we use undersampling technique to generate balanced training and test datasets. We repeated this method to generate ten different versions of training and test datasets. Each version of dataset had about 8000 songs with about 1000 songs for each genre. Lyrics of each genre were randomly split into two partitions: 80\% for training and 20\% for testing. Experimental results reported here are average across these ten datasets. We did not observe any significant variance in results across different instances of training and test datasets, indicating the robustness of the results.

\begin{figure} [bt]
\label{confusion}
\centering
\includegraphics[height=1.6in, width=4in]{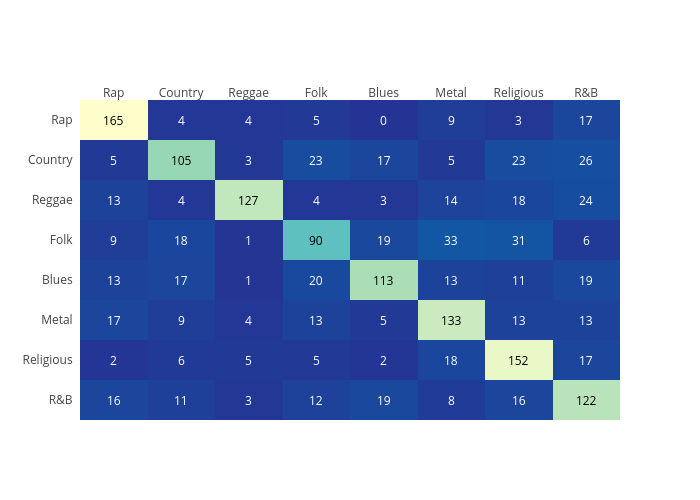}

\caption{Confusion Matrix for Genre. Rows:True Label, Columns:Predicted Label.}
\label{conf_genre}
\end{figure}

Distributed representation of lyrics and genres were jointly learned using doc2vec model~\cite{pmlr-v32-le14}. This model gave eight genre vectors (a vector representation for each genre) and vector representation for each song in the training and test dataset. We experimented with vector dimensionality and found 300 as the optimal dimensionality for our task. Using this vector representation, we experimented with both traditional machine learning models ( SVM, KNN, Random Forest and Genre Vectors) and deep learning models (CNN, GRU, and Bidirectional GRU) for genre prediction task. Please refer to Table ~\ref{table:genre}. For the KNN model, the genre of a test instance was determined based on genres of K nearest neighbors in the training dataset. Nearest neighbors were determined using cosine similarity. We tried three parameter values for K: 10, 25, and 50. However, there was no significant difference in results. For Genre Vector model, the genre of a test instance was determined based on the cosine similarity of test instance with vectors obtained for each genre. We can observe that Rap is the easiest genre to predict as rap songs have a distinctive vocabulary. The Folk genre is the most difficult to identify. For each genre, the worst performing model is the KNN, indicating that the local neighborhood of a test instance is not the best indicator of the genre. On average, Genre Vector model performs the best.

Please refer to Figure 3. This figure represents confusion matrix for one version of training dataset using Genre Vector model. Each row of the matrix sums to around 200 as the number of instances in test dataset per genre were around 200. We can notice that confusion relationships are asymmetric. We say that a genre X is confused with genre Y if the genre prediction model identifies many songs of genre X  as having genre Y. For example, observe the row corresponding to the Folk genre. It is mainly confused with the Religious genre as about 16\% of Folk songs are identified as Religious. However for the Religious genre, Folk does not appear as one of the top confused genres. Similarly, genre Reggae is most confused with R\&B. However, R\&B is least confused with Reggae.

\section{Popularity Prediction}
\label{section:popularity}
Only a subset of songs had user ratings data available with ratings ranging from 1 to 5 \cite{Fell2014LyricsbasedAA}. For two genres: Folk and Blues, we did not get popularity data for enough number of songs. For popularity prediction task, the number of genres was thus reduced to six. Number of songs per genre are: Metal(15254), Country(2640), Religious(3296), Rap(19774), R\&B(6144), and Reggae(294). Songs of each genre were randomly partitioned into two disjoint sets: 80\% for training and 20\% for testing. To ensure robustness of results, we performed experiments on ten such versions of the dataset. Experimental results reported here are average across ten runs. We model popularity prediction as a binary classification problem. For each genre, we divided songs into two categories: low popularity (rated 1, 2, or 3) and high popularity(rated 4 or 5). The number of songs included in each class were balanced to avoid any over fitting of model.

Considering the distinctive nature of each genre, we built a separate model per genre for popularity prediction. For each genre using the doc2vec model, we generated two popularity vectors (one each for low and high popularity) and vector representation for each song in training and testing dataset. Similar to the genre prediction task, we experimented with seven prediction models. Please refer to Table~\ref{popularity}. We can observe that Deep Learning based models perform better than other models. However, for every genre, the gap between the best and worst model has narrowed down as compared to the genre prediction task. 

\begin{table*}[bt]
\label{table:popularity}
\centering
\caption{F1-Scores for Popularity Prediction. Highest value for each genre is in bold.}
\label{popularity}
\makebox[\textwidth][c]{
\begin{tabular}{|c|c|c|c|c|c|c|}
\hline

\backslashbox{ Model }{ Genre }  & Country  &Metal       & Rap                          & Reggae                       & Religious                    & R\&B                         \\ 
\hline

SVM   &0.6238  & 0.6756           & 0.7301                   & 0.7539                            & 0.5681                  & 0.6342             \\
\hline

KNN     &0.5871     &0.6351      & 0.7071                & 0.7713                 & 0.5387             & 0.5814          \\
\hline

Random Forest   &0.6201    &0.6683     & 0.7176               & 0.7647                 & 0.5635           & 0.646          \\
\hline

Popularity Vector & 0.6180   &\textbf{0.6776}  & 0.7663    & 0.7820         & 0.5886     & 0.6401 \\
\hline
CNN   &\textbf{0.632}    &0.6717       & 0.7652             & \textbf{0.8011}             & \textbf{0.5933}                   &\textbf{0.6661}          \\
\hline

GRU   &0.5801  &0.6479           & 0.7505                    & 0.5187                            & 0.5661                  & 0.6434             \\
\hline

Bi-GRU  &0.6037    &0.6581       & \textbf{0.7684}             & 0.6613              & 0.5517                   & 0.5886          \\
\hline

\end{tabular}}
\end{table*}

\section{Improving Word Vectors with Lyrics}
\label{section:word analogy}

A large text corpus such as Wikipedia is necessary to train distributed representation of words. Lyrics are a poetic creation that requires significant creativity. Knowledge encoded in them can be utilized by training distributed representation of words. For this task, we used our entire dataset of 400K songs. Using the word2vec model, we generated four sets of word vectors. The four training datasets were: Lyrics only ($D1$, 470 MB), Complete Wikipedia ($D2$, 13 GB), Sampled Wikipedia ($D3$, 470 MB), and Lyrics combined with Wikipedia ($D4$, 13.47 GB). For dataset $D3$, we randomly sampled pages from Wikipedia till we collected dataset of a size comparable to our Lyrics dataset. For dataset $D3$, we created ten such sampled versions of Wikipedia. Results given here for $D3$ are average across ten such datasets.

\begin{table*}[bt]
\centering
\caption{Results of Word Analogy tasks. Highest value for each task is in bold.}
\label{table:word analogy}
\begin{tabular}{|c|c|c|c|c|}
\hline
  Tasks & Lyrics   & Wikipedia  &Sampled Wiki& Lyrics+Wiki             \\
\hline
1) capital-common-countries   & 10.95 & 87.75 & \textbf{89.43} &  87.94        \\
\hline
2) capital-world             & 07.26 &\textbf{90.35} & 79.25 &  90.00          \\
\hline
3) currency             & 02.94 &\textbf{05.56} & 01.85 &  \textbf{05.56}         \\
\hline
4) city-in-state               & 07.87 &66.55 & 61.71 &  \textbf{66.73}          \\
\hline
5) family             &81.05 & \textbf{94.74} & 82.82 &  94.15    \\
\hline
6) gram1-adjective-to-adverb  & 08.86 & 35.71 & 25.53 &  \textbf{36.51}         \\
\hline
7) gram2-opposite &19.88 & \textbf{51.47} & 33.46 &  51.10 
\\
\hline
8) gram3-comparative & 83.56 &\textbf{91.18} & 80.66 & 90.00 
\\
\hline
9) gram4-superlative &53.33 &75.72 &59.49 &\textbf{77.83} 
\\
\hline
10) gram5-present-participle &74.32 &73.33 &60.82 &\textbf{75.81} 
\\
\hline
11) gram6-nationality-adjective &06.29 &97.01 &92.60 &\textbf{97.08} 
\\
\hline
12) gram7-past-tense &54.44 &68.07 &65.47 &\textbf{69.00} 
\\
\hline
13) gram8-plural &73.19 &87.60 &76.81 &\textbf{89.52}
\\
\hline
14) gram9-plural-verbs &60.00 &71.85 &63.32 &\textbf{72.62} 
\\
\hline
Overall Across All Tasks &50.33 &75.71 &66.6 &\textbf{78.11} 
\\
\hline
\end{tabular}
\end{table*}

To compare these four sets of word embeddings, we used 14 tasks of word analogy tests proposed by Mikolov~\cite{DBLP:journals/corr/abs-1301-3781}. Please refer to Table~\ref{table:word analogy}. Each cell in the table represents accuracy (in percentage) of a particular word vector set for a particular word analogy task. First five tasks in the table consist of finding a related pair of words. These can be grouped as semantic tests. Next nine tasks (6 to 14) check syntactic properties of word vectors using various grammar related tests. These can be grouped as syntactic tests. 

By sheer size, we expect $D2$ to beat our dataset $D1$. However, we can observe that for tasks 5, 8, 12, 13, and 14 $D1$ gives results comparable to $D2$. For task 10, $D1$ is able to beat $D2$ despite the significant size difference. Datasets $D3$ and $D1$ are comparable in size. For task 10, $D1$ significantly outperforms $D3$. For all other tasks, the performance gap between $D3$ and $D1$ is reduced noticeably. We can observe that $D1$ performs better on syntactic tests than semantic tests. However, the main takeaway from this experiment is that dataset $D4$ performs the best for a majority of the tasks. Also, $D4$ is the best performing dataset overall. These results indicate that lyrics can be used in conjunction with large text corpus to further improve distributed representation of words.

\section{Related Work}
\label{section:relwork}
Existing works that have used lyrics for genre and popularity prediction can be partitioned into two categories.  First, that use lyrics in augmentation with acoustic features of the song \cite{McKay2010EvaluatingTG,Hu:2012:GCM:2348283.2348480} and second, that do not use acoustic features \cite{LoganIEEE,Mayer08,6204985,Fell2014LyricsbasedAA}. However, all of them represent lyrics using either handcrafted features or bag-of-words models. Identifying features manually requires intricate knowledge of music, and such features vary with the underlying dataset. Mikolov and Le have shown that distributed representation of words and documents is superior to bag-of-words models \cite{NIPS2013_5021,pmlr-v32-le14}. To the best of our knowledge, this is the first work that capitalizes on such representation of lyrics for genre and popularity prediction. However, our results cannot be directly compared with existing works as datasets, set of genres, the definition of popularity, and distribution of target classes are not identical. Still, our results stand in contrast with existing works that have concluded that lyrics alone are a weak indicator of genre and popularity. These works report significantly low performance of lyrics for genre prediction task. For example, Rauber et al. report an accuracy of 34\% \cite{Mayer08},  Doraisamy et al. report an accuracy of 40\% \cite{6204985}, McKay et al. report an accuracy of 43\% \cite{McKay2010EvaluatingTG}, and Hu et al. reported accuracy of abysmal 19\% \cite{Hu:2012:GCM:2348283.2348480}. The accuracy of our method is around 63\%.

\section{Conclusion and Future Work}
\label{section:conclusion}
This work has demonstrated that using distributed representation; lyrics can serve as a good indicator of genre and popularity. Lyrics can also be useful to improve distributed representation of words. Deep Learning based models can deliver better results if larger training datasets are available. Our method can be easily integrated with recent music mining algorithms that use an ensemble of lyrical, audio, and social features.

\bibliographystyle{abbrv}
\bibliography{lyrics}

\begin{thebibliography}{10}

\bibitem{midia17}
Lyrics take centre stage in streaming music, a midia research white paper,
  2017.
\newblock
  \url{https://www.nielsen.com/us/en/insights/reports/2018/2017-music-us-year-end-report.html}.

\bibitem{nielson17}
Nielsen 2017 u.s. music year-end report.
\newblock
  \url{https://www.midiaresearch.com/app/uploads/2018/01/Lyrics-Take-Centre-Stage-In-Streaming-%E2%80%93-LyricFind-Report.pdf}.

\bibitem{LoganIEEE}
P.~M. B.~Logan, A.~Kositsky.
\newblock Semantic analysis of song lyrics.
\newblock {\em IEEE International Conference on Multimedia and Expo (ICME)},
  pages 159--168, Jun 2004.

\bibitem{Fell2014LyricsbasedAA}
M.~Fell and C.~Sporleder.
\newblock Lyrics-based analysis and classification of music.
\newblock In {\em COLING}, 2014.

\bibitem{Hu:2012:GCM:2348283.2348480}
Y.~Hu and M.~Ogihara.
\newblock Genre classification for million song dataset using confidence-based
  classifiers combination.
\newblock In {\em ACM SIGIR International Conference on Research and
  Development in Information Retrieval}, pages 1083--1084, 2012.

\bibitem{pmlr-v32-le14}
Q.~Le and T.~Mikolov.
\newblock Distributed representations of sentences and documents.
\newblock In {\em International Conference on Machine Learning}, pages
  1188--1196, 2014.

\bibitem{McKay2010EvaluatingTG}
C.~McKay, J.~A. Burgoyne, J.~Hockman, J.~B.~L. Smith, G.~Vigliensoni, and
  I.~Fujinaga.
\newblock Evaluating the genre classification performance of lyrical features
  relative to audio, symbolic and cultural features.
\newblock In {\em ISMIR}, 2010.

\bibitem{DBLP:journals/corr/abs-1301-3781}
T.~Mikolov, K.~Chen, G.~Corrado, and J.~Dean.
\newblock Efficient estimation of word representations in vector space.
\newblock {\em CoRR}, abs/1301.3781, 2013.

\bibitem{NIPS2013_5021}
T.~Mikolov, I.~Sutskever, K.~Chen, G.~S. Corrado, and J.~Dean.
\newblock Distributed representations of words and phrases and their
  compositionality.
\newblock In {\em Advances in Neural Information Processing Systems}, pages
  3111--3119. 2013.

\bibitem{Mayer08}
A.~R. R.~Mayer, R.~Neumayer.
\newblock Rhyme and style features for musical genre classification by song
  lyrics.
\newblock {\em International Conference on Music Information Retrieval
  (ISMIR)}, pages 337--342, Jun 2008.

\bibitem{RudolfACM}
A.~R. Rudolf~Mayer, Robert~Neumayer.
\newblock Combination of audio and lyrics features for genre classification in
  digital audio collections.
\newblock {\em In Proceedings of the 16th ACM international conference on
  Multimedia}, pages 159--168, Oct. 2008.

\bibitem{CAMP12}
S.~D. T.~C.~Ying and L.~N. Abdullah.
\newblock Genre and mood classification using lyric features.
\newblock {\em Information Retrieval and Knowledge (CAMP), 2012 International
  Conference on. IEEE}, Mar. 2012.

\bibitem{6204985}
T.~C. Ying, S.~Doraisamy, and L.~N. Abdullah.
\newblock Genre and mood classification using lyric features.
\newblock In {\em International Conference on Information Retrieval Knowledge
  Management}, pages 260--263, 2012.

\bibitem{DBLP:journals/cim/YoungHPC18}
T.~Young, D.~Hazarika, S.~Poria, and E.~Cambria.
\newblock Recent trends in deep learning based natural language processing
  [review article].
\newblock {\em {IEEE} Comp. Int. Mag.}, 13(3):55--75, 2018.

\end{thebibliography}

\end{document}